\documentclass[pra,aps,twocolumn,showpacs,superscriptaddress,floatfix,amsmath,amssymb,nofootinbib]{revtex4}
\usepackage{amsfonts}
\usepackage{graphicx}
\newcommand{\beq}{\begin{equation}}
\newcommand{\eeq}{\end{equation}}

\newcommand{\beqa}{\begin{eqnarray}}
\newcommand{\eeqa}{\end{eqnarray}}
\newcommand{\beqax}{\begin{eqnarray*}}
\newcommand{\eeqax}{\end{eqnarray*}}

\def\ra{\rangle}
\def\la{\langle}
\begin{document}
\title{Hamiltonian engineering via invariants and dynamical algebra}
\author{E. Torrontegui}
\affiliation{Institute of Chemistry, The Hebrew University, Jerusalem 91904, Israel}
\affiliation{Departamento de Qu\'{\i}mica F\'{\i}sica, Universidad del Pa\'{\i}s Vasco UPV/EHU, 
Apdo. 644, Bilbao, Spain}
\author{S. Mart\'{\i}nez-Garaot}
\affiliation{Departamento de Qu\'{\i}mica F\'{\i}sica, Universidad del Pa\'{\i}s Vasco UPV/EHU, 
Apdo. 644, Bilbao, Spain}
\author{J. G. Muga}
\affiliation{Departamento de Qu\'{\i}mica F\'{\i}sica, Universidad del Pa\'{\i}s Vasco UPV/EHU, 
Apdo. 644, Bilbao, Spain}
\affiliation{Department of Physics, Shanghai University, 200444 Shanghai, People's Republic of China}
\begin{abstract}
%
%
%
We use the dynamical algebra of a quantum   
system and its dynamical invariants to inverse engineer feasible Hamiltonians 
for implementing shortcuts to adiabaticity. These are speeded up processes that 
end up with the same populations than slow, adiabatic ones.    
As application examples we design families of shortcut Hamiltonians   
that  drive two and a three-level systems between initial and
final configurations imposing physically motivated constraints 
on the terms (generators) allowed in the Hamiltonian. 
\end{abstract}  	
\maketitle
\section{Introduction}
The current  development of ``shortcuts to adiabaticity'' to speed up  adiabatic, slow processes
in different fields (trap expansions \cite{prl_Xi1, OCTexpan, schaff, 3d}, atom or ion transport \cite{David, Calarco, transport, BECtransport, OCT, ions}, internal state control 
\cite{Rice03, Rice05, Rice08, Berry2009, PRL2, Oliver, NHSara, Yue_prl}, wavepacket splitting \cite{S07, S09a, S09b, MNProc, FFgeneral, FFsplitting, multi}, many-body state engineering 
\cite{adolfo1, nat_adolfo, bjj1, bjj2,mbad,mbka}, optics \cite{optics}, cooling methods \cite{on1,on2,Wu}, 
or cooling cycles and quantum engines \cite{Salamon09, energy,EPL11, adolfo3}), raises a number of practical and fundamental questions (see \cite{review} for a recent review).  
An important one is to generate alternative shortcuts when the,    
generally time-dependent, Hamiltonian that speeds up the slow process 
is difficult or impossible to realize in the laboratory \cite{hosc,Sara_prl,mbad,mbka,review,Kazu}.
Typically the difficulties are related to specific terms that cannot be implemented. 
Several examples showed that the symmetry of the 
Hamiltonian is instrumental in designing feasible alternative Hamiltonians (and shortcuts) that keep 
the same population dynamics in some basis or at least the same final populations \cite{Sara_prl,review,Kazu}. 
However, a systematic symmetry-based approach to inverse engineering the Hamiltonian, 
given the desired dynamics and specific constraints imposed on its structure, was lacking. 
In this paper we provide  basic elements for such an approach and 
set  the inverse problem from a (Lewis and Riesenfeld \cite{LR})   dynamical invariant to the Hamiltonian, 
assuming that both are members of a dynamical Lie algebra, 
a vector space spanned by operators (generators) closed under commutation.  
Dynamical invariants correspond to operators whose expectation values remain constant for states evolving 
with the associated Hamiltonian (they may or may not commute with it). 
Invariants that belong to the dynamical Lie algebra of the Hamiltonian 
have been used to solve the dynamics, or to calculate geometric phases
\cite{Lieinv,KM,KP,MKN}.  
Since the dynamical invariants contain information of the system evolution (any density operator describing its evolution is 
a dynamical invariant), they have  also 
been used  to design shortcuts  to adiabaticity 
\cite{prl_Xi1, OCTexpan, schaff, 3d, transport, BECtransport, OCT,ions, Yue_prl,multi, adolfo1, nat_adolfo, bjj1, bjj2, adolfo3, inva-berry, review}
as we shall do here, taking explicitly into account the dynamical algebra in the Hamiltonian construction.   

In Sec. \ref{general}  a construction method is provided to design families of Hamiltonians for a given invariant 
in the space spanned by their corresponding algebra. This method allows to impose constraints on the  
generators, such as making some of them zero.  
As well, boundary conditions for the invariant are specified at initial and final instants so that the Hamiltonian indeed 
drives the system along a shortcut to adiabaticity without final excitation. 
%
We work out  two examples that illustrate the construction algorithm.  
In Sec. \ref{example} we construct real Hamiltonians within the SU(2) algebra to drive a two-level atom state 
without using the Pauli matrix 
$\sigma_y$.  
Then we analyze in Sec. \ref{example2} a three-level system described by a 4-dimensional Lie algebra, with the goal  
of achieving fast ``insulator-superfluid'' transitions for two interacting bosons in two-wells, using only two feasible generators. 
Conclusions and open questions are summarized in Sec. \ref{Discussion and conclusions}. 
The Appendix \ref{gauss} provides an alternative approach, using Gaussian elimination, 
to the operational approach of Sec. \ref{general}.  
\section{General formalism}
\label{general}
Let us assume that the time-dependent Hamiltonian $H(t)$ describing a quantum system  
%
%
is given by a linear combination of Hermitian ``generators'' $T_a$,
\beq
\label{ham}
H(t)=\sum_{a=1}^{N}h_a(t)T_a,
\eeq
where the $h_a(t)$ are real time-dependent functions 
and the $T_a$ span a  
Lie algebra (dynamical algebra \cite{Lieinv}) 
\beq
\label{lie}
[T_b,T_c]=\sum_{a=1}^N\alpha_{abc}T_a,  
\eeq
where the $\alpha_{abc}$ are the ``structure constants''. Associated with the Hamiltonian
there are time-dependent Hermitian invariants of motion $I(t)$ that satisfy \cite{LR}
\beq
\label{def}
\frac{dI}{dt}\equiv\frac{\partial I(t)}{\partial t}-\frac{1}{i\hbar}[H(t),I(t)]=0.
\eeq
A wave function $|\Psi(t)\rangle$ which evolves 
with $H(t)$ 
can be expressed as a linear combination of invariant modes  \cite{LR}
\beq
\label{wave}
|\Psi(t)\rangle=\sum_{n}c_ne^{i\alpha_n}|\phi_n(t)\rangle,
\eeq
where the $c_n$ are constants, the phases $\alpha_n$ fulfill
\beq
\hbar\frac{d\alpha_n}{dt}=\langle\phi_n(t)|i\hbar\frac{\partial}{\partial t}-H(t)|\phi_n(t)\rangle,
\eeq
and the eigenvectors of $I(t)$,  $|\phi_n(t)\ra$,  are assumed to form a complete set and satisfy 
\beq
I(t)|\phi_n(t)\rangle=\lambda_n|\phi_n(t)\rangle,
\eeq
$\lambda_n$ being the constant eigenvalues. 

If the invariant is a member of the dynamical algebra, it can be written as
\beq
\label{inv}
I(t)=\sum_{a=1}^{N}f_a(t)T_a,
\eeq
where $f_a(t)$ are real, time-dependent functions. Note that some $f_a$ or some $h_a$ may be zero. 
Replacing  Eqs. (\ref{ham}) and (\ref{inv})
into Eq. (\ref{def}), and using Eq. (\ref{lie}), the  functions $h_a(t)$ and $f_a(t)$ satisfy \cite{Lieinv,maamache}
\beq
\label{fund}
\dot f_a(t)-\frac{1}{i\hbar}\sum_{b=1}^{N}\sum_{c=1}^{N}\alpha_{abc}h_{b}(t)f_c(t)=0,  
\eeq
where the dot means time derivative.
Usually these coupled equations are interpreted as a linear system of ordinary differential 
equations for $f_a(t)$ when the $h_a(t)$ components of the Hamiltonian are known \cite{Lieinv,KM,KP,MKN,maamache}.
Instead, we put forward here a different, inverse perspective, and consider them an algebraic system to be solved for 
the $h_a(t)$, when the $f_a(t)$ are given.    
Defining the $N\times N$ matrix $\mathcal{A}$ as 
\beq
\label{A}
\mathcal{A}_{ab}(t)=\frac{1}{i\hbar}\sum_{c=1}^{N}\alpha_{abc}f_c(t),
\eeq
Eq. (\ref{fund}) can be written as
\beqa
\label{fun2}
\dot f_a(t)&=&\sum_{b=1}^{N}\mathcal{A}_{ab}(t)h_b(t),
\\
\label{compact}
&{\rm or}&  |\mathbf{\dot f}\rangle=\mathcal{A}|\mathbf{h}\rangle, 
\eeqa
where the kets are defined 
in terms of the components of each generator, for example,  
\beq
\bf{|}\mathbf{f}{\bf\ra}=\left(
\begin{array}{c}
f_1\\
f_2\\
...\\
f_N
\end{array}
\right).
\eeq
In this vector space we may naturally refer to $|\bf{h}\ra$ and $|\bf{f}\ra$ as the Hamiltonian and the invariant, respectively. 
Even though the context should avoid any confusion with the  vectors $|\Psi(t)\ra$ defined before in the state-vector space, 
the difference is nevertheless emphasized by the bold-face notation. 
There are  many Hamiltonians for a given invariant \cite{inva-berry} and   
we cannot generally invert Eq. (\ref{compact}) as $|\mathbf{h}\rangle=\mathcal{A}^{-1}|\mathbf{\dot f}\rangle$ to get $|\mathbf{h}\ra$.  
This means that det($\mathcal{A}$)$=0$, so at least one of the eigenvalues  $a^{(i)}(t)$ of the $\mathcal{A}$ matrix vanishes.  
To find a valid $|\mathbf{h}\ra$ in this case one may use Gauss elimination, as shown in the Appendix. 
Alternatively we shall  follow here a more compact and intuitive operational approach. 
The $\mathcal{A}$ matrix may be generally non-Hermitian.  
It  has $N$ right eigenvectors $\{|\mathbf{a}^{(i)}(t)\rangle\}$, $i=1,2,\dots,N$, \cite{NH_0, NHSara},
\beq
\mathcal{A}(t)|\mathbf{a}^{(i)}(t)\rangle=a^{(i)}(t)|\mathbf{a}^{(i)}(t)\rangle,
\eeq
and biorthonormal partners $\{|{\hat{\mathbf{a}}}^{(i)}(t)\rangle\}$,
\beq
\mathcal{A}^{\dag}(t)|\hat{\mathbf{a}}^{(i)}(t)\rangle=\big({a^{(i)}}(t)\big)^{\!*}|{\hat{\mathbf{a}}}^{(i)}(t)\rangle,
\eeq
where the asterisk means complex-conjugate and the dagger denotes the adjoint. 
These eigenvectors are normalized as 
\beq
\label{norma}
\langle{\hat{\mathbf{a}}}^{(i)}(t)|\mathbf{a}^{(j)}(t)\rangle=\delta_{ij}, 
\eeq
where  bras are defined as  $\la \mathbf{a}|=(a_1^*, a_2^*,...a_N^*)$, 
and the scalar product  as $\la \mathbf{a}|\mathbf{b}\ra=a_1^* b_1+a_2^*b_2+...+a_N^*b_N$.
They satisfy closure relations
\beq
\sum_{i=1}^{N}|\hat{\mathbf{a}}^{(i)}(t)\rangle\langle \mathbf{a}^{(i)}(t)|
=\sum_{i=1}^{N}| \mathbf{a}^{(i)}(t)\rangle\langle \hat{\mathbf{a}}^{(i)}(t)|=\mathbb{I}_N.
\eeq
We can thus write the operator $\mathcal{A}(t)$ and its adjoint as
\beqa
\label{op}
\mathcal{A}(t)&=&\sum_{i=1}^N|\mathbf{a}^{(i)}(t)\rangle a^{(i)}(t)\langle\hat{\mathbf{a}}^{(i)}(t)|,
\nonumber\\
\mathcal{A}^{\dag}(t)&=&\sum_{i=1}^N|\hat{\mathbf{a}}^{(i)}(t)\rangle \big({a}^{(i)}(t)\big)^{\!*}\langle \mathbf{a}^{(i)}(t)|.
\eeqa
Let us define the null-subspace projector $\mathcal{Q}$ of $\mathcal{A}$ associated with the $a^{(i)}(t)=0$ eigenvalue,
as $\mathcal{Q}=\sum_{i=1}^{Q}|\mathbf{a}^{(i)}(t)\rangle\langle\hat{\mathbf{a}}^{(i)}(t)|$, 
and the complementary projector $\mathcal{P}=\sum_{i=1}^{P}|\mathbf{a}^{(i)}(t)\rangle\langle\hat{\mathbf{a}}^{(i)}(t)|$. 
We have that  $\mathcal{P}+\mathcal{Q}=\mathbb{I}_N$ and $P+Q=N$. Note as well 
that $\mathcal{P}$ and $\mathcal{Q}$ commute with $\mathcal{A}$ and  the relations $\mathcal{P}^2=\mathcal{P}$, $\mathcal{Q}^2=\mathcal{Q}$, 
and $\mathcal{QA}=0$.  
To solve $|\mathbf{\dot f}\rangle=\mathcal{A}|\mathbf{h}\rangle$ for $|\mathbf{h}\ra$, we use Eq. (\ref{op}) 
and 
%
%
project it first onto the $\mathcal{P}$ subspace,
\beqa
\mathcal{P}|\mathbf{\dot f}\rangle&=&\sum_{i=1}^N\sum_{j=1}^P|\mathbf{a}^{(j)}\rangle\langle\hat{\mathbf{a}}^{(j)}|\mathbf{a}^{(i)}\rangle {a}^{(i)}\langle
\hat{\mathbf{a}}^{(i)}|\mathbf{h}\rangle,
\nonumber\\
%
%
&=&
\sum_{j=1}^P|\mathbf{a}^{(j)}\rangle a^{(j)}\langle\hat{\mathbf{a}}^{(j)}|\mathbf{h}\rangle.
\eeqa
Since here all $a^{(j)}(t)\neq0$ we can invert the expression,
%
%
%
\beq
\label{sol}
\sum_{i=1}^P|\mathbf{a}^{(i)}\rangle {a^{(i)}}^{-1}\langle\hat{\mathbf{a}}^{(i)}|\mathbf{\dot f}\rangle
=\sum_{i=1}^P|\mathbf{a}^{(i)}\rangle\langle\hat{\mathbf{a}}^{(i)}|\mathbf{h}\rangle,
\eeq
so the $\mathcal{P}$ part of the solution is given by 
\beq
\label{sol2}
\mathcal{P}|\mathbf{h}\rangle=\mathcal{B} |\mathbf{\dot f}\rangle,
\eeq
where ${\mathcal{B}}=\mathcal{P}\mathcal{B}\mathcal{P}=\mathcal{P} \mathcal{B}= \mathcal{B}\mathcal{P}=\sum_{i=1}^P|\mathbf{a}^{(i)}\rangle {a^{(i)}}^{-1}\langle\hat{\mathbf{a}}^{(i)}|$ is a  pseudoinverse matrix of $\mathcal{A}$, as  $\mathcal{ABA=A}$.  This relation implies $P$ equations 
for the $\{h_j\}$, $\{f_j\}$ and $\{\dot{f}_j\}$. 
If  instead $|\mathbf{\dot f}\rangle=\mathcal{A}|\mathbf{h}\rangle$ is projected onto the null subspace 
we get 
%
%
%
\beq
\label{cond}
\mathcal{Q}|\mathbf{\dot f}\rangle=\sum_{j=1}^Q|\mathbf{a}^{(j)}\rangle a^{(j)}\langle\hat{\mathbf{a}}^{(j)}|\mathbf{h}\rangle=0,
\eeq
because now  all $a^{(j)}(t)=0$. 
%
%
%
%
This relation  implies  the existence of multiple solutions for $|\mathbf{h}\ra$, 
and $Q$ conditions $\la \hat{\mathbf{a}}^{(j)}|\mathbf{\dot{f}}\ra=0$ which involve  $\{f_i\}$ and their time derivatives.    
We can add any arbitrary part $\mathcal{Q}|\mathbf{h}\rangle$ to Eq. (\ref{sol2}) so that all Hamiltonians of the form 
\beq
\label{fundamental}
|\mathbf{h}\rangle={\mathcal{B}}|\mathbf{\dot f}\rangle+\mathcal{Q}|\mathbf{h}\rangle,
\eeq
where $\mathcal{Q}|\mathbf{h}\rangle$ is a completely arbitrary vector in the null subspace, are thus 
compatible with the invariant $I$.  
This is one of the fundamental equations of this paper. Due to the freedom to choose $\mathcal{Q}|\mathbf{h}\ra$ 
(we may construct it as  $\mathcal{Q}|\mathbf{b}\ra$, where $|\mathbf{b}\ra$ is arbitrary), we can 
change the Hamiltonian to make it realizable. In addition, the invariant itself may be modified.  

When inverse engineering shortcuts to adiabaticity, the Hamiltonian is usually given at initial and final times. 
In general the invariant $I$ (equivalently $|\mathbf{f}(t)\ra$) is chosen   
to drive,  through its eigenvectors,  the initial states of the Hamiltonian $H(0)$ to the states of the final $H(t_f)$ \cite{LR, prl_Xi1, transport}. 
This is ensured by imposing at the boundary times $t_b=0, t_f$,  the ``frictionless conditions'' $[H(t_b),I(t_b)]=0$ \cite{prl_Xi1}.
For Eqs. (\ref{ham}) and (\ref{inv}) these boundary conditions can be reformulated as
\beq
\sum_{a,b,c}^{N}\alpha_{abc}h_b(t_b)f_c(t_b)T_a=0. 
\eeq
Since the $T_a$ generators are independent the coefficients must satisfy
\beq
\sum_{b,c}^{N}\alpha_{abc}h_b(t_b)f_c(t_b)=0, \quad a=1,\dots,N,\quad t_b=0,t_f,
\eeq
or more compactly 
\beq
\label{flc}
\mathcal{A}(t_b)|\mathbf{h}(t_b)\ra=0,\quad t_b=0,t_f,
\eeq
a second fundamental result. 
Note that the choice of $\mathcal{Q}|\mathbf{h}\ra$  does not affect this condition, but $|\mathbf{f}\ra$ must be chosen to fulfill it.     
At the boundary times Eq. (\ref{flc}) imposes $N$ conditions, and, if the $N$ values of the $\{h_j(t_b)\}$ are given, 
the $P+Q$ equations  in Eqs. (\ref{sol2}) and (\ref{cond}) will fix the values of $\{f_j(t_b)\}$ and $\{\dot{f}_j(t_b)\}$. 
At intermediate times the Hamiltonian and invariant components can be designed    
subjected to the $N$ equations in Eqs. (\ref{sol2}) and (\ref{cond}) and to the boundary conditions.
This leaves open different inverse engineering 
possibilities: in general the Hamiltonian is first fixed partially, i.e., imposing the time dependence (or vanishing) of 
some $r<N$ components. Fixing the invariant time dependence consistently with the boundary conditions and the imposed 
Hamiltonian constraints, finally leads to equations that give the form of the remaining $N-r$ Hamiltonian components.  
The following sections illustrate these steps and concepts explicitly.  
\section{Example 1:  SU(2) Lie algebra}
\label{example}
%
%
%
%
%
%
%
%
%
%
%
Let us consider the SU(2)  
algebra ($N=3$) spanned by $\{T_1,T_2,T_3\}$ with commutation relations
\beq
\label{LieSU2}
[T_1,T_2]=iT_3,\; [T_2,T_3]=iT_1,\;[T_3,T_1]=iT_2.
\eeq
%
%
%
Equation (\ref{fun2}) becomes
\beq
\label{sisSU2}
\left(\begin{array}{c} 
\dot f_1 \\
\dot f_2   \\
\dot f_3 
\end{array} \right)=
\underbrace{\frac{1}{\hbar}\left(\begin{array}{ccc} 
0&   f_3 & -f_2  \\
-f_3 &   0& f_1 \\
f_2      &  -f_1 & 0 
\end{array} \right)}_{=\mathcal{A}}
\left(\begin{array}{c} 
h_1  \\
h_2\\
h_3
\end{array} \right).
\eeq
As $\mathcal{A}=-\mathcal{A}^{\dagger}$ is a real antisymmetric matrix and with  odd dimensionality, the
eigenvalues are conjugate pure imaginary pairs, and zero, whereas 
left and right eigenvectors are equal.
Explicitly the eigenvalues are $a^{(0)}=0$, $a^{(1)}=-i\sqrt{\gamma}/\hbar$, and $a^{(2)}=i\sqrt{\gamma}/\hbar$ 
(we have shifted down the 
superscripts by one with respect to the general formalism, here $j=0,1,2$,  so that the zero corresponds
to the zero eigenvalue),
with
\beq
\label{gamma}
\gamma=f_1^2+f_2^2+f_3^2,
\eeq
and  the eigenvectors 
\beqa
|a^{(0)}\rangle=|\hat a^{(0)}\rangle&=&\frac{1}{w_0}\left(
	\begin{array}{c}
	\frac{f_1}{f_3}\\
	\frac{f_2}{f_3}\\
	1
	\end{array} \right), \nonumber\\
|a^{(1)}\rangle=|\hat a^{(1)}\rangle&=&\frac{1}{w_1}\left(
	\begin{array}{c}
	\frac{-f_1f_3-if_2\sqrt{\gamma}}{\beta-\gamma}\\
	\frac{-f_1^2-f_3^2}{f_2f_3+if_1\sqrt{\gamma}}\\
	1
	\end{array} \right), \nonumber\\	
|a^{(2)}\rangle=|\hat a^{(2)}\rangle&=&\frac{1}{w_2}\left(
	\begin{array}{c}
	\frac{-f_1f_3+if_2\sqrt{\gamma}}{\beta-\gamma}\\
	\frac{-f_1^2-f_3^2}{f_2f_3-if_1\sqrt{\gamma}}\\
	1
	\end{array}\right),
\eeqa
where $w_2=w_1=\sqrt{2\gamma/(\beta-\gamma)}$, $w_0=\sqrt{\gamma/f_3^2}$, and 
\beq
\label{beta}
\beta=2(f_1^2+f_2^2)+f_3^2.
\eeq
The $\mathcal{P}$ and $\mathcal{Q}$ projectors are
\beq
\mathcal{P}=\mathbb{I}_3-\mathcal{Q}, \quad
\mathcal{Q}=\frac{1}{\gamma}{\left(\begin{array}{ccc} 
f_{1}^{2} &   f_1f_2 & f_1f_3  \\
f_1f_2 &   f_{2}^{2} & f_2f_3  \\
f_1f_3     &  f_2f_3 & f_{3}^{2} 
\end{array} \right),}
\eeq
whereas the pseudoinverse matrix $\mathcal{B}$, see Eq. (\ref{sol2}), is 
\beq
\mathcal{B}=\frac{\hbar}{\gamma}{\left(\begin{array}{ccc} 
0 &   -f_3 & f_2  \\
f_3 &   0 & -f_1  \\
-f_2     &  f_1 & 0
\end{array} \right).}
\eeq
%
\begin{figure}[t!]
\begin{center}
\includegraphics[width=4.2cm]{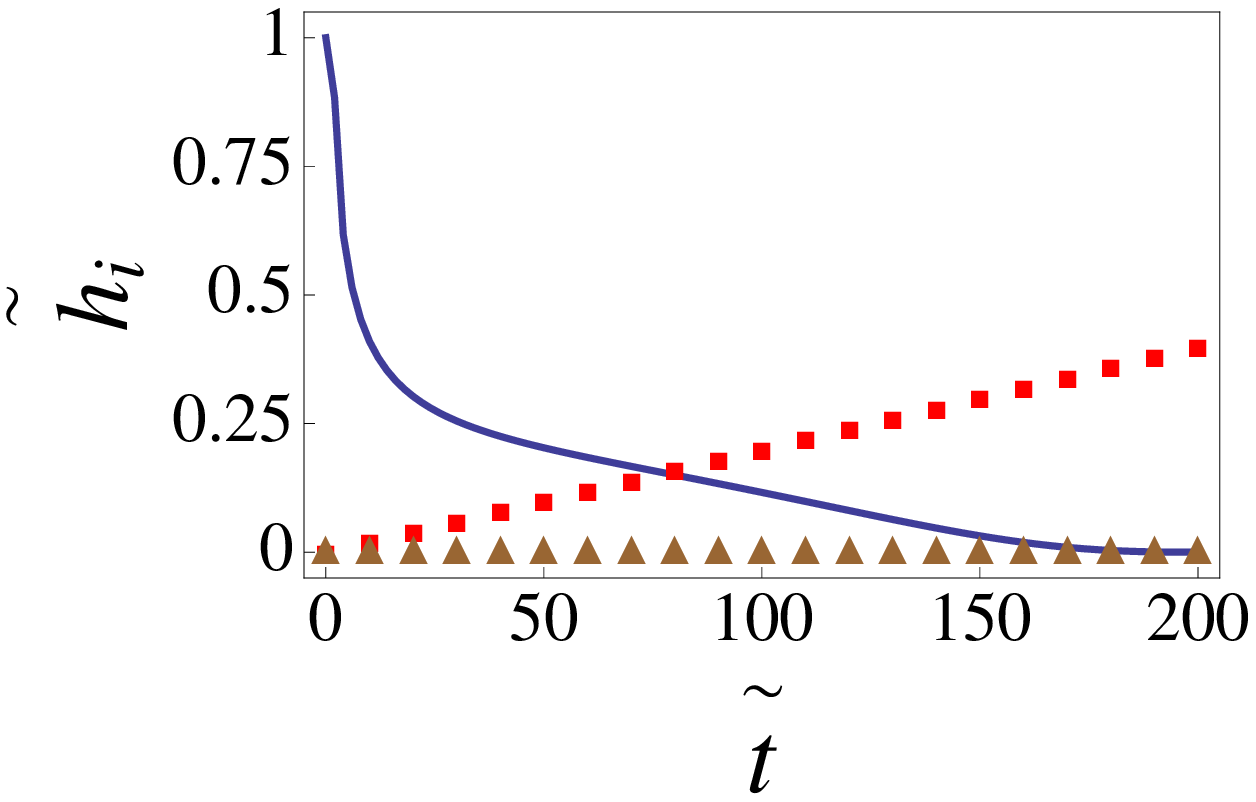}
\includegraphics[width=4.2cm]{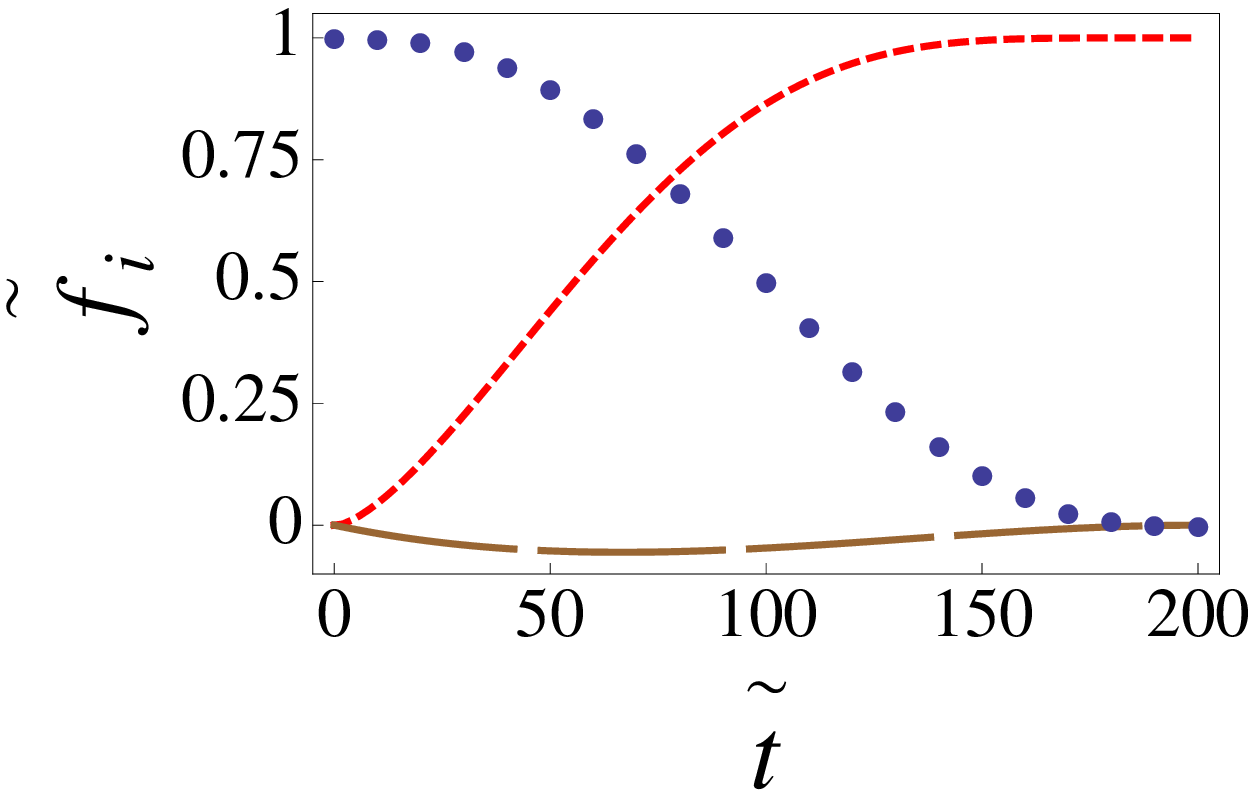}
\caption{(Color online) Hamiltonian (left) and invariant (right) coefficients versus time.
The imposed functions (symbols) are $h_1(t)$ (red squares), $h_2(t)$ (brown triangles),  and $f_3(t)$ (blue circles). 
The derived functions (lines) are $h_3(t)$ (blue solid), $f_1(t)$ (red short-dashed), and $f_2(t)$ (brown long-dashed).  
Parameter values: $\tilde h_1(t_f)=0.4$, $\tilde h_3(0)=1$,  and $\tilde t_f=200$.
 }
\label{f1}
\end{center}
\end{figure}
%
%
%
%
%
%
From Eq. (\ref{cond}) we
get the condition that the system has infinite solutions,
\beq
\left(\begin{array}{c} 
\frac{f_1(f_1\dot f_1+f_2\dot f_2+f_3\dot f_3)}{f_1^2+f_2^2+f_3^2}  \\
\frac{f_2(f_1\dot f_1+f_2\dot f_2+f_3\dot f_3)}{f_1^2+f_2^2+f_3^2}  \\
\frac{f_3(f_1\dot f_1+f_2\dot f_2+f_3\dot f_3)}{f_1^2+f_2^2+f_3^2}
\end{array} \right)=0. 
\eeq
As there is only one independent eigenvector for the null subspace, $Q=1$, 
this is solved by a single condition, 
\beq
\label{conder}
f_1\dot f_1+f_2\dot f_2+f_3\dot f_3=0,
\eeq
i.e., $\gamma=constant$, and the $f_i$ cannot be arbitrary independent functions.  
Equation (\ref{sol2}) becomes 
\beq
\mathcal{P}\left(\begin{array}{c} 
h_1 \\
h_2 \\
h_3
\end{array} \right)=\frac{\hbar}{\gamma}
\left(\begin{array}{c} 
f_2\dot f_3-f_3\dot f_2  \\
-f_1\dot f_3-f_3\dot f_1  \\
f_1\dot f_2-f_2\dot f_1
\end{array} \right), 
\eeq
and the general solution (\ref{fundamental}) takes the form 
\beq
\left(\!\begin{array}{c} 
h_1  \\
h_2  \\
h_3
\end{array}\! \right)\!=\!\frac{\hbar}{\gamma}
\left(\!\begin{array}{c} 
-\dot f_2f_3\!+\!f_2\dot f_3\!+\!f_1\la\mathbf{h}|\mathbf{f}\ra/\hbar  \\
\dot f_1f_3\!-\!f_1\dot f_3\!+\!f_2\la\mathbf{h}|\mathbf{f}\ra/\hbar  \\
-\dot f_1f_2\!+\!f_1\dot f_2\!+\!f_3\la\mathbf{h}|\mathbf{f}\ra/\hbar
\end{array}\! \right).
\eeq
%
%
%
%
Using Eq. (\ref{conder}),   
this gives the compact result 
\beq
\label{Hs1_sol}
h_i=-\hbar{\cal{E}}_{ijk}\frac{\dot f_j}{f_k}+\frac{f_i}{f_k}h_k,
\eeq
with all indices $i,\, j,\, k$ different. ${\cal{E}}_{ijk}$ is the Levy-Civita 
symbol (1 for even permutations of (123) and -1 for odd permutations)  
and $h_k(t)$ is considered a free function chosen 
for convenience. If we want to impose, for example, that one of the 
components of the Hamiltonian is zero, then we take that component to be 
$h_k$.    
%
%
%
%
%
%
%
%
\subsection{Two-level system}
To be more specific let us consider the following representation useful to 
describe, for example, a  two-level system in an external driving field,  
\beq
\!\!\!\!\!\!T_1=\frac{1}{2}\left(\begin{array}{cc} 
0 & 1 \\
1 & 0
\end{array} \right)\!,
T_2=\frac{1}{2}\left(\begin{array}{cc} 
0 & -i \\
i & 0
\end{array} \right)\!, 
T_3=\frac{1}{2}\left(\begin{array}{cc} 
1 & 0 \\
0& -1
\end{array} \right)\!. 
\eeq
%
We set as initial and final constraints the following Hamiltonians 
%
\beqa
\label{ini}
H(0)&=&h_1(0)T_1+h_3(0)T_3,
\\
%
%
\label{fin}
H(t_f)&=&h_1(t_f)T_1+h_3(t_f)T_3,
\eeqa
but in general $H(t)=h_1(t)T_1+h_2(t)T_2+h_3(t)T_3$. 
The frictionless conditions (\ref{flc}) for SU(2) read
\beqa
\label{fric}
f_i(t_b)h_j(t_b)-f_j(t_b)h_i(t_b)=0,\;\;  i>j.
\eeqa
Our aim is to find $H(t)$ so that the ground and excited states of $H(0)$ become ground and excited states of $H(t_f)$
in an arbitrary time $t_f$, up to phase factors,  in such a way that $h_2(t)=0$ $\forall t$. 
This condition is motivated by the difficulty to implement $T_2$ 
in some systems \cite{Oliver}.
Choosing $(i,j,k)=(1,3,2)$ in Eq. (\ref{Hs1_sol}), with  $h_2(t)=0$ and using $\gamma=constant=c_1$, we can express
$f_2$ and $f_1$ in terms of $f_3$, 
\beqa
\label{eqF}
f_2&=&\frac{\hbar\dot f_3}{h_1},
\nonumber\\
f_1&=&\sqrt{c_1-f_3^2-\frac{\hbar^2\dot f_3^2}{h_1^2}}.  
\eeqa
Substituting this in the other equation of Eq. (\ref{Hs1_sol}),  with $(i,j,k)=(3,1,2)$,  
\beq
\label{eqF1}
\ddot f_3-\frac{\dot h_1}{h_1}\dot f_3+\frac{h_1}{\hbar^2}\bigg(f_3h_1-h_3\sqrt{c_1-f_3^2-\frac{\hbar^2\dot f_3^2}{h_1^2}}\bigg)=0.
\eeq
%
%
%
%
%
%
%
\begin{figure}[t!]
\begin{center}
\includegraphics[width=4.2cm]{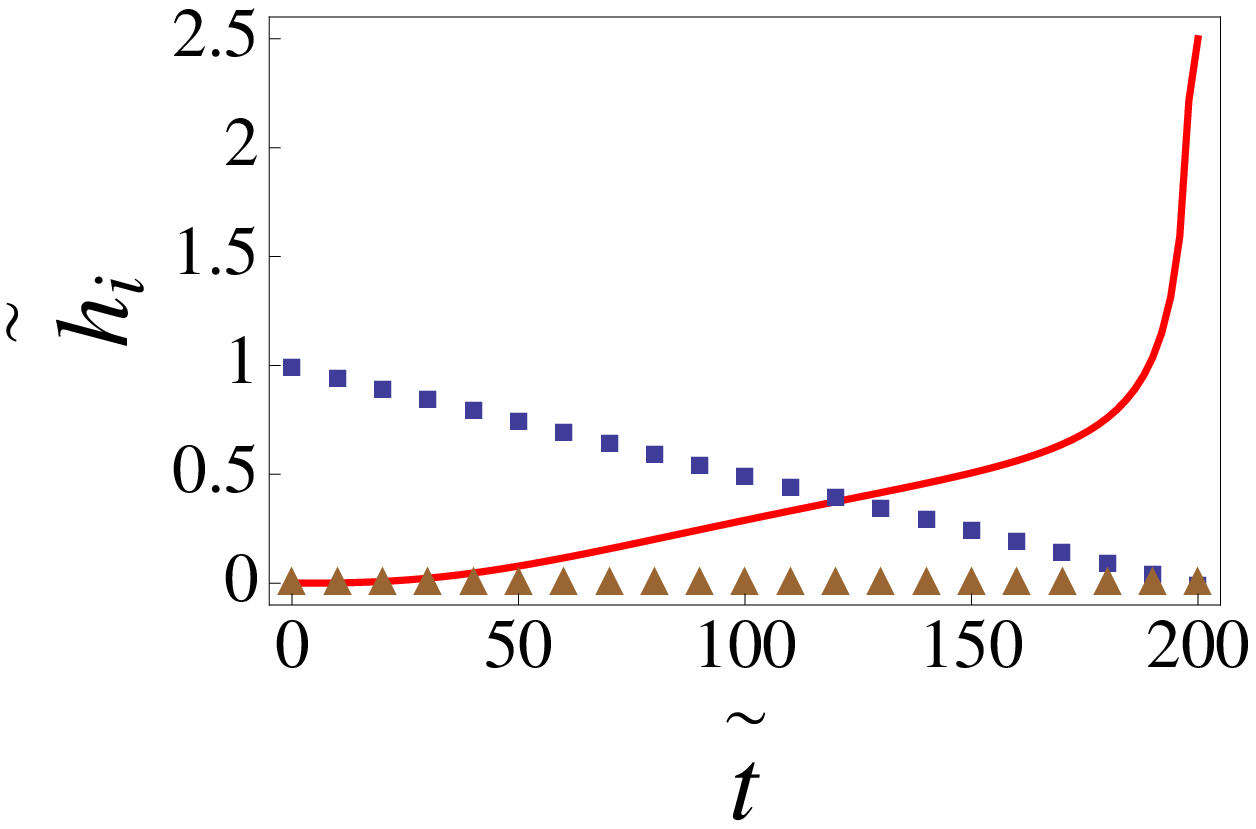}
\includegraphics[width=4.2cm]{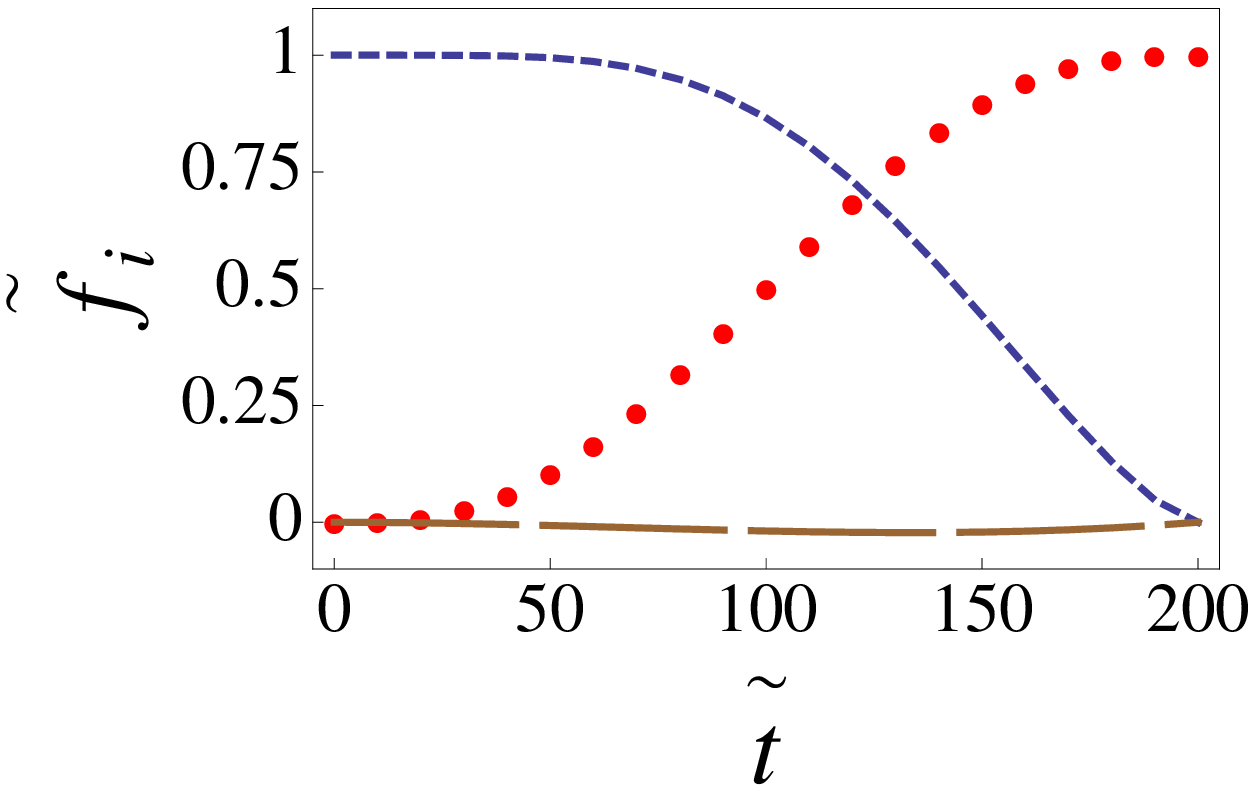}
\caption{(Color online) Hamiltonian (left) and invariant (right) coefficients versus time.
The imposed functions (symbols) are $h_2(t)$ (brown triangles), $h_3(t)$ (blue squares),  and $f_1(t)$ (red circles). 
The derived functions (lines) are $h_1(t)$ (red solid), $f_2(t)$ (brown long-dashed), and $f_3(t)$ (blue short-dashed).
Parameter values: $\tilde h_3(0)=1$, $\tilde h_1(t_f)=2.5$, and $\tilde t_f=200$.
 }
\label{f3}
\end{center}
\end{figure}
%
%
The frictionless conditions (\ref{fric}) for this case impose $f_2(t_b)=0$ and $h_3(t_b)/h_1(t_b)=f_3(t_b)/f_1(t_b)$ or, equivalently,
\beq
\label{co1}
f_3(t_b)=h_3(t_b)\sqrt{\frac{c_1}{h_1^2(t_b)+h_3^2(t_b)}},\quad \dot f_3(t_b)=0.
\eeq
In addition, from Eq. (\ref{eqF1}) at the boundary times $t_b$,
\beq
\label{co2}
\ddot f_3(t_b)=0.
\eeq
An example of  possible Hamiltonian engineering strategy is to impose first, in addition to $h_2(t)=0$, also the form of  $h_1(t)$. 
We then interpolate $f_3(t)$ (with a simple polynomial or following some more sophisticated approach, e. g. to optimize some variable) 
satisfying the boundary  conditions (\ref{co1},\ref{co2}) at the 
boundary times $t_b$, and solve for $h_3(t)$ in Eq. (\ref{eqF1}).
In the example of Fig. \ref{f1} the initial ground state of Eq. (\ref{ini}) with $h_1(0)=0$ is placed at the north pole
of the Bloch sphere and it is 
driven to the equator of the sphere ending as the ground state of Eq. (\ref{fin}),  
with $h_3(t_f)=0$. 
$h_3(t)$ is deduced assuming  $h_1(t)=h_1(t_f){t}/{t_f}$ and $f_3(t)=\sum_{i=0}^5b_it^i$, where
the $b_i$ coefficients are deduced from the boundary conditions (\ref{co1}) and (\ref{co2}). 
We set $c_1=h_1^2(0)+h_3^2(0)$ so that $H(0)=I(0)$.  
In this and the rest of figures we plot dimensionless variables  $\tilde t=t\sqrt{c_1}/\hbar$
and $\tilde E=E/\sqrt{c_1}$, and the invariant has been set with dimensions of energy. 
To find solutions with real functions the condition 
\beq
\label{condition}
f_3^2+\frac{\hbar^2\dot f_3^2}{h_1^2}\leq c_1\quad\forall t 
\eeq
must be satisfied, see 
Eqs. (\ref{eqF}) and (\ref{eqF1}). This sets a minimum final time $t_{f,m}$ that 
depends on the constant $c_1$ and the 
ansatz to interpolate $f_3(t)$. 
The (dimensionless) minimum time for the parameter values considered in Fig. \ref{f1} is $\tilde t_{f,m}\sim 165$.

Note that Eq. (\ref{eqF1}) is an algebraic equation for $h_3(t)$ and a differential equation for $h_1(t)$. 
Other option is to set $h_3(t)$ first and deduce $h_1(t)$ from Eq. (\ref{eqF1}), which is now a differential equation for this variable.
To solve instead for $h_1(t)$ algebraically, we may engineer $f_1(t)$ rather than $f_3(t)$, 
with an equation of the form of Eq. (\ref{eqF1}) that has the indices 1 and 3 swapped, see Fig. \ref{f3}. 
%
%
%
%
%
%
%
%
\section{Example 2: U3S3}
\label{example2}
As a second example consider the 4-dimensional ($N=4$) Lie algebra U3S3 \cite{MacCallum}, 
with basis   
$\{T_1,T_2,T_3,T_4\}$, where $\{T_1,T_2,T_3\}$ span a SU(2) subalgebra, see Eq. (\ref{LieSU2}), 
whereas the only non vanishing commutators of $T_4$ are  

\beq
\label{abelian}
[T_4,T_1]=iT_2, \quad [T_2,T_4]=iT_1, 
\eeq
Since $T_4-T_3$ commutes with any member of the algebra (it is an invariant in a Lie-algebraic sense),
this combination might appear as the natural 
fourth basis generator instead of $T_4$.  However, the use of $T_4$  is physically motivated by its natural occurrence in 
the system we shall deal with, namely, two interacting bosons in two wells \cite{Molmer}.  
The $\mathcal{A}$ matrix for our basis choice is 
\beq
\mathcal{A}=\frac{1}{\hbar}{\left(\begin{array}{cccc} 
0 &   -f_3 & f_2 & f_2  \\
f_3 &   0 & -f_1 & f_1 \\
-f_2     &  f_1 & 0 & 0 \\
-f_2     &  f_1 & 0 & 0
\end{array} \right),}
\eeq
with eigenvalues $a^{(0)}=a^{(1)}=0$, $a^{(2)}=-i\beta/\hbar$, and $a^{(3)}=i\beta/\hbar$. The left and right
eigenvectors are  
\beqa
|a^{(0)}\rangle=|\hat a^{(0)}\rangle&=&\frac{1}{\sqrt{\gamma}}\left(
	\begin{array}{c}
	f_1\\
	f_2\\
	0\\
	f_3
	\end{array} \right),	
\nonumber\\
|a^{(1)}\rangle=|\hat a^{(1)}\rangle&=&\frac{1}{\sqrt{\gamma\beta}}\left(
	\begin{array}{c}
	f_1f_3\\
	f_2f_3\\
	\gamma\\
	\beta-\gamma
	\end{array} \right), 
\nonumber\\		
|a^{(2)}\rangle=|\hat a^{(2)}\rangle&=&\frac{1}{\sqrt{2\beta}}\left(
	\begin{array}{c}
	\frac{-f_1f_3+if_2\sqrt{\beta}}{\beta-\gamma}\\
	\frac{2f_1^2+f_3^2}{-f_2f_3+if_1\sqrt{\beta}}\\
	\sqrt{\beta-\gamma}\\
	\sqrt{\beta-\gamma}
	\end{array}\right), 
\nonumber\\	
|a^{(3)}\rangle=|\hat a^{(3)}\rangle&=&\frac{1}{\sqrt{2\beta}}\left(
	\begin{array}{c}
	\frac{-f_1f_3-if_2\sqrt{\beta}}{\beta-\gamma}\\
	\frac{-f_2f_3+if_1\sqrt{\beta}}{\beta-\gamma}\\
	\sqrt{\beta-\gamma}\\
	\sqrt{\beta-\gamma}
	\end{array} \right),
\eeqa
where $\gamma$ and $\beta$ are defined as before, see Eqs. (\ref{gamma}) and (\ref{beta}). 
A novelty with respect to the previous example is that the zero eigenvalue is degenerate, so the null-subspace has dimension $Q=2$. 
Equation (\ref{cond}) sets now two conditions, 
one is the same condition as for SU(2), $\dot f_1f_1+\dot f_2f_2+\dot f_3f_3=0$, 
and the other one is $\dot f_3=\dot f_4$, so the system $|\mathbf{\dot f}\rangle=\mathcal{A}|\mathbf{h}\rangle$ has solution 
for $|\mathbf{h}\rangle$ if
\beqa
f_1^2+f_2^2+f_3^2&=&c_1\label{c1},
\\
f_4&=&f_3+c_2,
\label{c2}
\eeqa
with $c_1$ and $c_2$ constants (there are two independent $f_i$). 
We use now 
Eq. (\ref{fundamental}) as in the previous example
to get
\beqa
\label{ham_su2+1}
h_1&=&\frac{\hbar(\dot f_1f_1+\dot f_2f_2)+f_1f_3h_2}{f_2f_3},
\nonumber\\
h_4&=&\frac{\hbar\dot f_1+f_3h_2-f_2h_3}{f_2}.
\eeqa
Due to degeneracy of the null eigenvalue there are two free $h_i$, and we have chosen them to be $h_2$ and $h_3$ here. The frictionless
conditions $\mathcal{A}(t_b)|\mathbf{h}(t_b)\rangle=0$ become explicitly
\beqa
-f_3(t_b)h_2(t_b)+f_2(t_b)h_3(t_b)+f_2(t_b)h_4(t_b)&=&0, \nonumber\\
f_3(t_b)h_1(t_b)-f_1(t_b)h_3(t_b)+f_1(t_b)h_4(t_b)&=&0, \nonumber\\
-f_2(t_b)h_1(t_b)+f_1(t_b)h_2(t_b)&=&0. \label{fc}
\eeqa
\subsection{Two interacting bosons in a double well}
%
%
%
%
\begin{figure}[t!]
\begin{center}
\includegraphics[width=4.2cm]{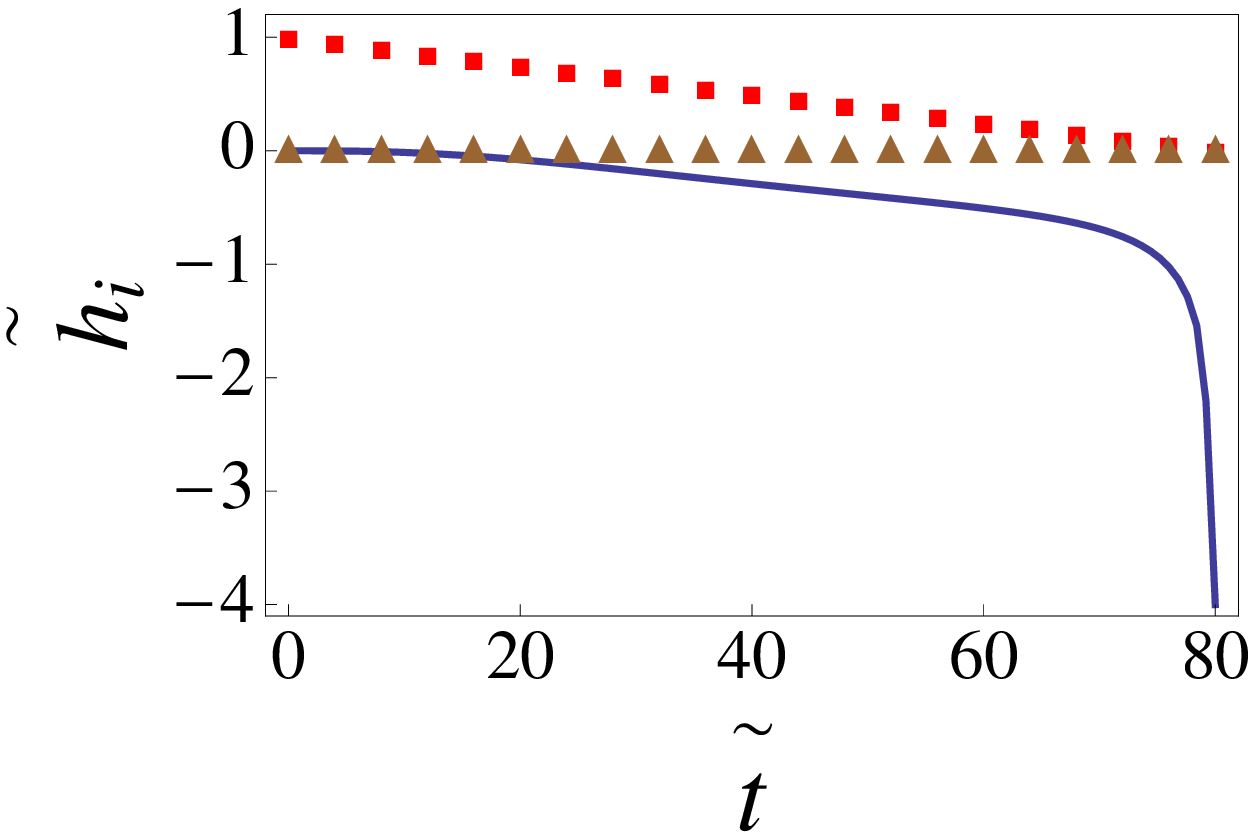}
\includegraphics[width=4.2cm]{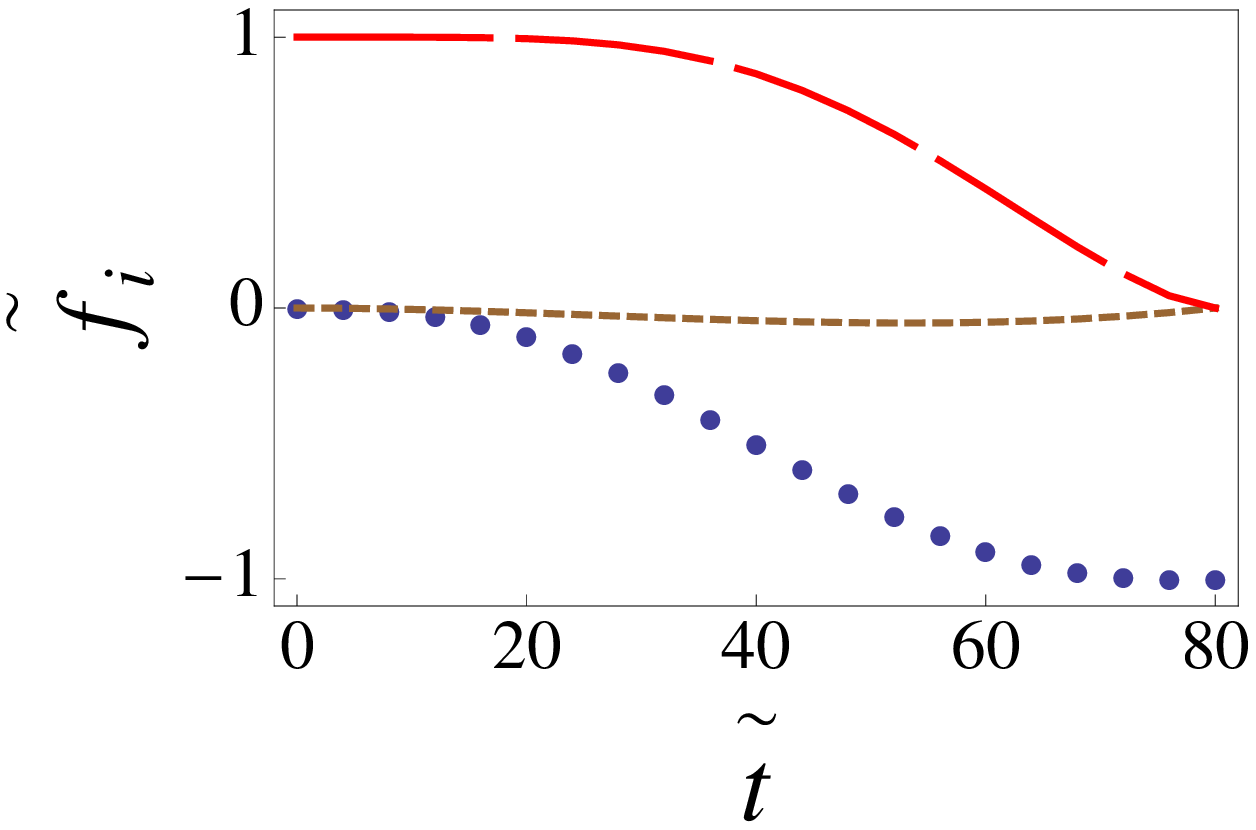}
\caption{(Color online) Hamiltonian (left) and invariant (right) coefficients versus time.
The imposed functions (symbols) are $h_2(t)$ (brown triangles), $h_3(t)$, (brown triangles), $h_4(t)$ (red squares), and $f_1(t)$ (blue circles). 
The derived functions (lines) are $h_1(t)$ 
(blue solid), $f_2(t)$ (brown, short-dashed),  $f_3(t)$ 
(red long-dashed), and $f_4(t)$ 
(red long-dashed). Parameter values: $\tilde h_1(t_f)=-4$, $\tilde h_4(0)=1$, and $\tilde t_f=80$.}
\label{f4}
\end{center}
\end{figure}
%
%
%
An interacting boson gas in a lattice potential can be described by the Bose-Hubbard model \cite{Fisher, Jaksch}.   
For two particles in two wells, the Hamiltonian in the occupation number basis $\{|n_{left}, n_{right}\ra\}$: 
$|2,0\rangle=\left ( \scriptsize{\begin{array} {rcccl} 1\\ 0\\0 \end{array}} \right)$, $|1,1\rangle=\left ( \scriptsize{\begin{array} {rcccl} 0\\ 1\\0 \end{array}} \right)$ and $|0,2\rangle=\left ( \scriptsize{\begin{array} {rcccl} 0\\ 0\\1 \end{array}} \right)$, is given by \cite{Molmer}
\beq
\label{H_0}
H_0=\left ( \begin{array}{ccc}
U & -\sqrt{2}J & 0\\
-\sqrt{2}J & 0 & -\sqrt{2}J \\
0 & -\sqrt{2}J & U
\end{array} \right)=UT_4-4JT_1,
\eeq
where $U$ gives the interparticle interaction energy and $J$ is a hopping constant, assumed to be controllable
functions of time, and 
\beq
\label{generators}
T_1=\frac{1}{2\sqrt{2}}\left ( \begin{array}{ccc}
0 & 1 & 0\\
1 & 0 & 1 \\
0 & 1 & 0
\end{array} \right), 
T_4=\left ( \begin{array}{ccc}
1 & 0 & 0\\
0 & 0 & 0 \\
0 & 0 & 1
\end{array} \right).
\eeq
We may close the algebra with two additional generators, 
\beq
\label{generators}
T_2=\frac{1}{2\sqrt{2}}\left ( \begin{array}{ccc}
0 & -i & 0\\
i & 0 & i \\
0 & -i & 0
\end{array} \right),  
T_3=\frac{1}{4}\left ( \begin{array}{ccc}
1 & 0 & 1\\
0 & -2 & 0 \\
1 & 0 & 1
\end{array} \right), 
\eeq
where the  $T_a$ satisfy the commutation 
relations given by Eqs. (\ref{LieSU2}) and (\ref{abelian}). 
Let us consider that at initial and final times the Hamiltonian of our system is
\beqa
H(0)&=&h_1(0)T_1+h_4(0)T_4, \label{h0}\\
H(t_f)&=&h_1(t_f)T_1+h_4(t_f)T_4,\label{hF}
\eeqa
and we want to drive without final excitation the ground state of $H(0)$ to $H(t_f)$. In \cite{Molmer} the shortcut-to-adiabaticity method 
followed requires the addition of the ``counterdiabatic term'' proportional to $T_2$, which is difficult to implement \cite{Molmer}. 
$T_3$ is also problematic so we   
shall engineer the Hamiltonian with Eq. (\ref{ham_su2+1}) imposing $h_2(t)=h_3(t)=0$ $\forall t$. Then using Eq. (\ref{c1}),
\beqa
\label{efes}
f_2&=&\frac{\hbar\dot f_1}{h_4}, \nonumber\\
f_3&=&\sqrt{c_1-f_1^2-\frac{\hbar^2\dot f_1^2}{h_4^2}},
\eeqa
where $f_1$ satisfies
\beq
\label{eqF1_b}
\ddot f_1-\frac{\dot h_4}{h_4}\dot f_1+\frac{h_4}{\hbar^2}\bigg(f_1h_4-h_1\sqrt{c_1-f_1^2-\frac{\hbar^2\dot f_1^2}{h_4^2}}\bigg)=0.
\eeq
%
 Additionally $f_4$ is given by Eq. (\ref{c2}). The frictionless conditions (\ref{fc}) for $h_2(t_b)=h_3(t_b)=0$ are
\beq
\label{co1_b}
f_1(t_b)=h_1(t_b)\sqrt{\frac{c_1}{h_1^2(t_b)+h_4^2(t_b)}},\quad \dot f_1(t_b)=0,
\eeq
and from Eq. (\ref{eqF1_b}), at the boundary times $t_b$,
\beq
\label{co2_b}
\ddot f_1(t_b)=0.
\eeq
Assuming that $h_4(t)$ is imposed,  Eq. (\ref{eqF1_b}) sets $h_1(t)$ to drive the initial ground state of $H(0)$ to $H(t_f)$ without
undesired terms. In Fig. \ref{f4} the Hamiltonian and the invariant components  
are plotted for a frictionless Mott-insulator to superfluid transition \cite{Molmer, Sof}.
The initial ground state of Eq. (\ref{h0}) with $h_1(0)=0$ corresponds to a Mott insulator and it evolves into the superfluid ground state of Eq.(\ref{hF}) with $h_4(t_f)=0$. 
We assume a linear variation of $h_4(t)=h_4(0)(1-{t}/{t_f})$ and $f_1(t)=\sum_{i=0}^5\bar{b}_it^i$, where
the $\bar{b}_i$ are deduced from the boundary conditions (\ref{co1_b}) and (\ref{co2_b}), $c_1=h_1^2(0)+h_4^2(0)$, and $c_2=0$.

Similarly to the previous example, if we impose the form of $h_1(t)$ instead of $h_4$, 
$h_4(t)$ can be deduced algebraically from Eq. (\ref{eqF1_b}) 
replacing $f_1$ by  $f_3$ and then swapping 4 and 1.

\section{Outlook}
\label{Discussion and conclusions}
We have worked out a framework to engineer time-dependent Hamiltonians and speed up adiabatic processes
making use of dynamical invariants and dynamical algebras. This is particularly useful to find shortcuts
free from Hamiltonian terms difficult to implement in practice.    
Explicit construction formulae allow us to fix some components of the Hamiltonian, to make them zero, for example, 
and get the time dependence of the remaining components.  

This work should be distinguished from a related method presented in  a companion paper \cite{Sof}. 
Both approaches share the use of Lie algebraic methods 
and the aim of constructing shortcuts. However the approach presented here is a systematic bottom-up inverse engineering method   
based on the relation between Hamiltonian and dynamical invariants. Instead, in \cite{Sof} the dynamical 
invariants do not play an explicit role. The starting point for the approach in \cite{Sof} is an existing, known shortcut;  
then, unitary transformations are carried out to generate alternative, 
(feasible or more convenient) shortcuts, as in \cite{Sara_prl}.     
The connection between the two approaches is left for a separate study. 

We also mention some fundamental questions worth investigating:  The type and structure 
of the algebra is expected to determine the inverse-engineering possibilities and limitations,
which are still little known. In particular the role of Lie-algebraic invariants (in contrast to dynamical invariants), 
or subalgebras should be examined \cite{casimir}, and quantum control theory \cite{qc}, which  overlaps in part with our objectives, 
may shed light on permissible or precluded operations. Adapting the current ideas to many-body 
systems is a further open question that may benefit from approaches based on restricting the action of the Hamiltonian to a 
subspace \cite{mbka}.   

The emphasis here has been on the cancellation of undesired Hamiltonian terms, but other applications of the 
proposed Hamiltonian engineering are possible. For example, to  optimize variables or 
robustness versus noise \cite{ruido1,ruido2}, transient energy,
and other relevant variables \cite{OCT}.   
Finally, the formalism proposed may be extended to open systems governed by dynamical equations 
formulated by closed Lie algebras \cite{Yair, refri, Tova, Amikam}.  
\acknowledgements{  
We are grateful to K. Takahashi and R. Kosloff for stimulating discussions.   
We acknowledge funding by Grants No. IT472-10 and No. FIS2009-12773-C02-01, and the UPV/EHU Program UFI 11/55.
E. T. is supported by the Basque Government postdoctoral program. S. M.-G. acknowledges a UPV/EHU fellowship.} 
\appendix
\section{Gauss elimination}
\label{gauss}
A way to solve the system (\ref{compact}) for $|\mathbf{h}(t)\ra$ 
is to use Gauss elimination. We consider explicitly the SU(2) group.  
The augmented matrix associated with the system in Eq. (\ref{sisSU2}) is
\beq
{\left(\begin{array}{cccc} 
-f_3 &   0 & f_1  &\hbar\dot f_2\\
0 &   f_3&- f_2 &\hbar\dot f_1\\
f_2      & -f_1 & 0 &\hbar\dot f_3
\end{array} \right)}.
\eeq
The essence of the method is to reduce the system to an equivalent one with the same solutions applying elementary operations. 
These are the multiplication of a row by a non-zero scalar,
the interchange of columns or rows, and the addition to a row of the multiple of a different one. 
In a first step $(f_2/f_3)$ times the first row is added to the third one
\beq
{\left(\begin{array}{cccc} 
-f_3 &   0 & f_1  &\hbar\dot f_2\\
0 &   f_3&- f_2 &\hbar\dot f_1\\
0      & -f_1 & \frac{f_1f_2}{f_3} &\frac{\hbar f_2\dot f_2}{f_3}+\hbar\dot f_3
\end{array} \right)}.
\eeq
Finally $(f_1/f_3)$ times the second row is added to the third producing a lower triangular matrix,
\beq
{\left(\begin{array}{cccc} 
-f_3 &   0 & f_1  &\hbar\dot f_2\\
0 &   f_3 & -f_2  &\hbar\dot f_1\\
0     &  0 & 0 &\frac{\hbar(f_1\dot f_1+f_2\dot f_2+f_3\dot f_3)}{f_3}
\end{array} \right)}.
\eeq
This system is compatible and has infinite solutions if $f_1\dot f_1+f_2\dot f_2+f_3\dot f_3=0$ 
or, equivalently,  $f_1^2+f_2^2+f_3^2=c_1$. The solutions satisfy
\beqa
\hbar\dot f_1&=&f_3h_2-f_2h_3,
\nonumber\\
\hbar\dot f_2&=&-f_3h_1+f_1h_3,
\eeqa
from which Eq. (\ref{Hs1_sol}) follows. 
%
%
%
%
%
%
%
%
%
%
%
%
%
%
%

%

%
\end{document}